\documentstyle[12pt,aps,prc,epsfig]{revtex}

\begin{document}
\title{Critical Behaviour and Universality in
Gravitational Collapse of a Charged Scalar Field}
\author{Shahar Hod and Tsvi Piran  \\
The Racah Institute for Physics, \\
The Hebrew University, Jerusalem, 91904 Israel \\
}
\date{28 July 1996}
\maketitle

\begin{abstract}
We summarize results from a study of spherically symmetric collapse of
a {\it charged} (complex) massless scalar-field \cite{Hod}. We present
an analytic argument which conjecture the generalization of the
mass-scaling relation and echoing phenomena, originally discovered by
Choptuik, for the {\it charged} case. Furthermore, we study the
behaviour of the self-similar critical solution under {\it external}
perturbations -- addition of a cosmological constant $\Lambda$ and a
charge-conjugation $e$. Finally, we study the scaling-relation of the
black-hole charge. Using an analytic argument we conjecture that
black-holes of infinitesimal mass are neutral or obey the relation
$Q_{BH} \ll M_{BH}$. We verify our predictions with numerical results.
\end{abstract}

\section{INTRODUCTION}

Gravitational collapse is one of the most interesting phenomena in
general relativity. The dynamics of a spherically-symmetric massless
scalar-field coupled to general relativity has two kinds of possible
end states.  Either the scalar field eventually dissipates away
leaving spacetime flat or a black-hole forms. Numerical simulations of
this model problem \cite{Choptuik} have revealed a very interesting
phenomena -- a kind of critical behaviour which is a feature of
supercritical initial conditions very close to the critical case $p =
p^*$ ($p$ is a parameter which characterizes the strength of the
initial configuration, and $p^*$ is the threshold value). More
precisely, Choptuik found a power-law dependence of the black-hole
mass on critical separation $p - p^*$ of the form:
\begin{equation}\label{eq1}
M_{BH}\propto  \left\{ {\matrix{0&{p\le p^*}\cr
{\left( {p-p^*} \right)^\beta }&{p>p^*}\cr
}} \right.
\end{equation}

Subsequently the same type of critical behaviour has been observed for
other collapsing fields: the collapse of axisymmetric gravitational
wave packets \cite{Abrahams}, the collapse of spherically symmetric
radiative fluids \cite{Coleman}.  In all these model problems the
critical exponent $\beta$ turned out to be close to the value
originally found by Choptuik $\beta \approx 0.37$, suggesting a
universal behaviour. However, Maison \cite{Maison} has shown that for
fluid collapse models with an equation of state given by $p = k\rho$
the critical exponent strongly depends on the parameter $k$.

The second key feature of Choptuik's results is the {\it universality}
of the precisely critical $(p = p^*)$ evolution at the threshold of
black-hole formation -- it was found that the critical solution has a
discrete self-similar behaviour (discrete echoing with a period
$\Delta$).

In this paper we summarize results from a study of spherically
symmetric collapse of a {\it charged} (complex) massless scalar field.
The main question we consider is whether it is possible to generalize
Choptuik's results for the general {\it charged} case \cite{Hod}. This
generalization is not trivial since the introduction of
charge-conjugation destroys the invariance of the evolution equations
under the rescaling $u \to au, r \to ar$, where $a$ is an arbitrary
positive constant. This invariance characterizes the system in the
special neutral case and is crucial for the self-similarity of the
critical evolution. This lack of invariance hints that
charge-conjugation might destroy the phenomena in the general
situation.

Using a semi-quantitative arguments, which is based upon different
behaviour of the mass and charge under the rescaling $r \to ar$, we
conjecture that the significance and influence of the charge decreases
during the evolution and that the critical behaviour
appears. Following this we provide a numerical evidence which confirms
the generalization of the mass-scaling relation and echoing phenomena
for the {\it charged} situation.

So far, the mass-scaling relation of the black-hole, and the influence
of perturbations on the critical evolution itself, have been studied
in context of internal perturbations in the initial conditions, such
as a deviation of the field's amplitude from the critical one.  It is
of interest to study the behaviour of the critical solution under {\it
external} perturbations. We consider here the influence of the charge
$e$ as such a perturbation. We also consider the effect of another
external parameter, the cosmological constant $\Lambda$.

The plan of the paper is as follows. In Sec. II we describe the
evolution equations. In Sec. III we describe the algorithm and
numerical methods. In Sec. IV we describe our discretization and error
analysis. In Sec. V we describe our {\it theoretical} predictions and
compare them with our {\it numerical} results. Sections V.A and V.B
establish both qualitatively and quantitatively as well, the
generalization of the mass-scaling relation and echoing phenomena for
the {\it charged} case. In Sec. V.C we study the behaviour of the
critical evolution under two such external perturbations: addition of
a cosmological-constant $\Lambda$ to the critical solution and an
addition of a charge-conjugation $e$ to critical initial-conditions of
a complex neutral scalar-field. We also study the possibility of
forming black-holes from subcritical $(p < p^*)$ initial conditions
using external perturbations. In Sec. V.D we study the charge-mass
relation for infinitesimal black-holes.  We show that for $p - p^* \to
0$ the black-hole charge tends to zero more rapidly than its
mass. From this we conclude that black-holes of infinitesimal mass,
which can be created from near--critical evolutions, are neutral, or
obey the relation $Q_{BH} \ll M_{BH}$. Our numerical results confirm
this conjecture. We end in Sec. VI with a brief summary and
conclusions.

\section{THE EVOLUTION EQUATIONS}

We consider a spherically symmetric charged scalar field $\phi$. This
is a combination of two real scalar fields $\phi_1, \phi_2$, which are
combined into a complex one $\phi = \phi_1 + i\phi_2$. The
electromagnetic field is described by the potential $A$, which is
defined up to the addition of a gradient of a scalar function. The
electromagnetic field tensor $F$ is defined as $2dA$, i.e. $F_{ ab} =
2A_{[b;a]}$.

The total Lagrangian of the scalar field and electromagnetic field is
\cite{Hawking}:
\begin{equation}\label{eq2}
L = - {{1}\over{2}} (\phi_{;a} + ieA_a \phi)g^{ ab}
(\phi^*_{;b} - ieA_b  \phi^*) - {{1}\over{16 \pi}} F_{ ab}
F_{cd} g^{ac} g^{bd}~.
\end{equation}
Where $e$ is a constant and $\phi^*$ is the complex conjugate of
$\phi$.

Varying $\phi, \phi^*$ and $A_a$ independently, one obtains
\cite{Hawking}:
\begin{equation}\label{eq3}
\phi_{; ab}g^{ ab} + ieA_ag^{ ab} (2 \phi_{;b} + ieA_b
\phi) +
ieA_{a;b}g^{ ab}
\phi = 0~,                                       
\end{equation}
and its complex conjugate, and:
\begin{equation}\label{eq4}
{{1}\over{4\pi}}F_{ ab;c}g^{bc} - ie \phi (\phi^*_{;a} -
ieA_a \phi^* ) + ie \phi^* (\phi_{;a} + ieA_a \phi) = 0~.
\end{equation}
We express the metric of a spherically symmetric
spacetime in the form \cite{Christadoulou,Goldran}:
\begin{equation}\label{eq5}
ds^2 = -g(u,r) \bar g(u,r)du^2 - 2g(u,r)dudr + r^2d
\Omega^2~.
\end{equation}
The radial coordinate $r$ is a geometric quantity which
directly measures proper surface area, and $u$ is a
retarded time null coordinate. Here $d \Omega^2$ is the
two-sphere metric.

Because of the spherical symmetry, only the radial
electric field $F^{01} = -F^{10}$ is nonvanishing. This
choice satisfies Maxwell's equation:
\begin{equation}\label{eq6}
F_{[ ab;c]} = 0~.
\end{equation}
We introduce the auxiliary field $h$:
\begin{equation}\label{eq7}
\phi = \bar h \equiv {{1}\over{r}}\int\limits_0^rhdr~, 
\end{equation}
Using the radial component of Eq. (\ref{eq4}), we express
the charge contained within the sphere of radius $r$, at a
retarded time $u$, as:
\begin{equation}\label{eq8}
Q(u,r) = 4\pi i e \int\limits_0^r{r(\bar h^*h - \bar h
h^*)dr}~. 
\end{equation}
and the potential as:
\begin{equation}\label{eq9}
A_0 = \int\limits_0^r{{{Q}\over{r^2}}gdr}
\end{equation}
The energy-momentum tensor of the charged scalar-field is
\cite{Hawking}:
\begin{eqnarray}\label{eq10}
T_{ ab}={1 \over 2}\left( {\phi _{;a}\phi _{;b}^*+\phi
_{;a}^*\phi _{;b}} \right) & + & {1 \over 2}\left( {-\phi
_{;a}ieA_b\phi ^*+\phi _{;b}^*ieA_a\phi +\phi
_{;a}^*ieA_b\phi -\phi _{;b}ieA_a\phi ^*} \right) \nonumber \\
& + & {1 \over {4\pi
}}F_{ac}F_{bd}g^{cd}+e^2A_aA_b\phi \phi ^*+Lg_{ ab}~. 
\end{eqnarray} 
The nontrivial Einstein equations are:
\begin{equation}\label{eq11}
G_{rr}:~~{{2}\over{r}}{{g_{,r}}\over{g}}=8 \pi \bar h_{,r}
\bar h^*_{,r}~;
\end{equation}
\begin{equation}\label{eq12}
G_{ur}:~~{{1}\over{r^2}}\bar g \left[{{g}\over{\bar g}} +
r {{\bar g}\over{g}} \left({{g}\over{\bar g}}
\right)_{,r} -1 \right] = 8 \pi \left({{1}\over{2}} \bar g
\bar h_{,r} \bar h^*_{,r} + {{1}\over{8 \pi}} g
{{Q^2}\over{r^4}} \right)~.
\end{equation}
Regularity at the origin requires $g(u,0) = \bar g
(u,0)$. The boundary condition $h(u,0) = \bar h(u,0)$
forces us to integrate the equations outward, and impose
the normalization $g(u,0) = \bar g(u,0) = 1$, which
corresponds to selecting the time coordinate as the 
proper time on the $r = 0$ central world line.

The solution at a given $r$ depends only on the solution
at $r' < r$. We integrate Eq. (\ref{eq11}) and obtain:
\begin{equation}\label{eq13}
g(u,r) = \exp \left[ 4\pi \int\limits^r_0 {{{(h
- \bar h) (h^* - \bar h^*)}\over{r}}dr} \right]~,
\end{equation}
Using Eqs. (\ref{eq11}) and (\ref{eq12}), we obtain after
integration:
\begin{equation}\label{eq14}
\bar g(u,r) = {{1}\over{r}} \int\limits_0^r{\left(
1 - {{Q^2}\over{r^2}}\right)g dr}~.
\end{equation}
In terms of the variable $h$, the wave-equation Eq.
(\ref{eq3}) takes the form 
\begin{equation}\label{eq15}
Dh\equiv h_{,u}-{1 \over 2}\bar gh_{,r}={1 \over
{2r}}\left( {g-\bar g} \right)\left( {h-\bar h}
\right)-{{Q^2} \over {2r^3}}\left( {h-\bar h}
\right)g-{{ieQ} \over {2r}}g\bar h-iehA_0~.
\end{equation}
Using the characteristic method, we convert the
scalar-field evolution-equation Eq. (\ref{eq15}) into a
pair of coupled differential equations:
\begin{equation}\label{eq16}
{{dh}\over{du}}={1 \over
{2r}}\left( {g-\bar g} \right)\left( {h-\bar h}
\right)-{{Q^2} \over {2r^3}}\left( {h-\bar h}
\right)g-{{ieQ} \over {2r}}g\bar h-iehA_0~;
\end{equation}
\begin{equation}\label{eq17}
{{dr}\over{du}} = -{{1}\over{2}} \bar g~.
\end{equation}
We solve these equations together with the integral
equations (\ref{eq7}), (\ref{eq13}), (\ref{eq14}),
(\ref{eq8}) and (\ref{eq9}). The mass contained within the
sphere of radius $r$ at a retarded time $u$, is:
\begin{equation}\label{eq18}
M(u,r) \equiv {{r}\over{2}} \left( 1 -
{{\bar g}\over{g}} + {{Q^2}\over{r^2}} \right)~. 
\end{equation}
Using Eqs. (\ref{eq13}) and (\ref{eq14}), we express $M$
as:
\begin{equation}\label{eq19}
M(u,r) = \int\limits^r_0{\left[ 2 \pi
{{\bar g}\over{g}} \left(h - \bar h\right)
\left(h^* - \bar h^*\right) + {{1}\over{2}}
{{Q^2}\over{r^2}}\right]dr} + {{1}\over{2}}
{{Q^2}\over{r}}~.
\end{equation}

\section{ALGORITHM AND NUMERICAL METHODS.}

A numerical simulation of the special uncharged case was
first performed by Goldwirth and Piran \cite{Goldran}.
Gundlach, Price and Pullin \cite{Gundlach} used a version
of this algorithm to study the scaling behavior of the
mass of the black-hole for the special uncharged case. We
have used a version of the algorithm of Refs.
\cite{Goldran,Garfinkle}, for the neutral case, and we
have generalized it for the charged case. However, the
methods used in  Refs \cite{Goldran,Gundlach} are not
accurate enough for a treatment of the critical solution
itself, because each successive echo appears on spatial
and temporal scales a factor $e^{2\Delta} \approx 31$
finer than its predecessor.

The main improvements of our version closely resembles
that of Ref. \cite{Garfinkle}: Taylor expansion of the physical
quantities near the origin, conservation of the number of
grid points using interpolation and choosing the
outermost grid point to be the ingoing light ray that hits
the zero mass singularity of the critical solution $(p =
p^*)$. 

To solve numerically Eqs. (\ref{eq16}) and (\ref{eq17}) we
define a radial grid $r_n$, where $n = 1, \dots, N$. We
should emphasize that $\bar h$ is a complex field: $\bar
h = \bar h_1 + i \bar h_2$, and obtain a set of $3N$
coupled differential equations. 
\begin{eqnarray}\label{eq20}
{{dh_{1n}} \over {du}} = {1 \over
{2r_n}}\left( {g_n-\bar g_n} \right)\left( {h_{1n}-\bar
h_{1n}} \right) & - & {{Q^2_n} \over {2r^3_n}}\left(
{h_{1n}-\bar h_{1n}} \right)g_n \nonumber \\
& + & {{eQ_n} \over
{2r_n}}g_n\bar h_{2n}+eh_{2n}A_{0n}~;
\end{eqnarray}
\begin{eqnarray}\label{eq21}
{{dh_{2n}} \over {du}} = {1 \over
{2r_n}}\left( {g_n-\bar g_n} \right)\left( {h_{2n}-\bar
h_{2n}} \right) & - & {{Q^2_n} \over {2r^3_n}}\left(
{h_{2n}-\bar h_{2n}} \right)g_n \nonumber \\
& - &{{eQ_n} \over {2r_n}}g_n\bar
h_{1n}-eh_{1n}A_{0n}~;
\end{eqnarray}
\begin{equation}\label{eq22}
{{dr_n}\over{du}} = -{{1}\over{2}} \bar g_n~.
\end{equation}
where $g$ and $h$ satisfy the boundary conditions
\begin{equation}\label{eq23}
\bar g_1 = g_1 = 1~~;~~~\bar h_{11} = h_{11}~~;~~~\bar h_{21} = h_{21}
\end{equation}

The initial data for the Einstein-scalar-Maxwell equations
is just the value of $h$ on the initial data surface, $u =
0$, the value of the parameter $e$, and a numerical
choice of the initial position in $r$ of each ingoing
null lines of the grid. The algorithm proceeds as
follows: first we integrate along $r$ for a fixed
$u$ and find in succession the quantities $\bar h, g, Q,
A_0$ and $\bar g$.

We use Eqs. (\ref{eq7}), (\ref{eq13}), (\ref{eq8}),
(\ref{eq9}) and (\ref{eq14}) respectively in order to
evaluate these quantities.

The integration is carried out using a three-point
Simpson method for unequally spaced abscissas (Even if
the grid is evenly spaced initially, it will not remain
so during the evolution \cite{Goldran}). We next use Eqs.
(\ref{eq20}-\ref{eq22}), to evolve $h$ and $r$ one time
step forward. We solve the $3N$ ordinary differential
equations using the fifth-order Runge-Kutta method
\cite{Press}.
This process is iterated as many times as necessary: i.e.
until either the field disperses or a charged black-hole
forms.

The time step $\Delta u$ is determined so that in each
step the change in $r_n$ is less half the distance
between it and the null trajectory $r_{n-1}$, i.e.:
\begin{equation}\label{eq24}
\Delta u < {{r_n - r_{n-1}}\over{\bar g_n}}~.
\end{equation}
Once a null trajectory arrives at the origin $r=0$, it
bounces and disperses along $u =const.$ to infinity.
The grid-point is therefore lost when the light ray hits
the origin.

If for a given shell $M > Q$ at some time then it is
possible that
\begin{equation}\label{eq25}
\bar e^{2 \beta}(r,u) = 1 - {{2M(r,u)}\over{r}} +
{{Q^2(r,u)}\over{r^2}}
\end{equation}
will vanish. We identify the formation of a black hole
when there is an $r$ value that satisfies
\begin{equation}\label{eq26}   
r_\pm = M(r_\pm,u) \pm (M^2(r_\pm,u) -
Q^2(r_\pm,u))^{1/2}~. 
\end{equation}
In this case $r_\pm$ are the horizons of the shell.

As we approach the stage when a black-hole forms $\bar
g_n$ becomes infinite. In our algorithm this means that
the step size $\Delta u \to 0$. The numerical approach to
the horizon is stopped eventually by an overflow of
$\bar g$ or underflow of $\Delta u$. We can, still,
estimate where and when a black-hole  horizon appears
from the condition $r_+/r \to 1$.

In \cite{Gundlach} this method was used to study the
scaling behavior of the mass of the uncharged black-hole.
However, this method is not accurate enough for a
treatment of the critical solution itself. The main
improvements of our version for the charged case closely
resembles that of Ref. \cite{Garfinkle} where this method was
used to study the critical solution for the special
neutral case. We will now describe shortly the sources of
inaccuracy and the methods that we use to overcome them.

The first source of an inaccuracy arises from the
fact that the expressions for the quantities $\bar h, g,
\bar g, A_0$ and $M$ contains an explicit factor of $1/r$
which diverges to the origin. We overcome this by using a
Taylor expansion of $h$ in $r$:
\begin{equation}\label{eq27}
h = h^{(0)} + h^{(1)}r + h^{(2)}r^2 + O(r^3)~.
\end{equation}
Then we expand the quantities $\bar  h, Q, g, \bar g,
A_0$ and $M$ in $r$, where the expansion coefficients are
all functions of $h^{(0)}, h^{(0)*}, h^{(1)}, h^{(1)*},
h^{(2)}$ and $h^{(2)*}$. Thus, in order to treat the
solution near the origin one needs to find only these 
coefficients. This is done by fitting the first three
values of $h$ to a second-order polynome: i.e.  we  solve
equation (\ref{eq27}) for $r_1, r_2$ and $r_3$ to obtain
$h^{(0)}, h^{(1)}$ and $h^{(2)}$. We then use these
coefficients to evaluate the values of $\bar h,
Q, g, \bar g$ and $A_0$ for the first two values of $r$.
The values of these quantities for other values of $r$ is
then determined using a three-point Simpson method for
unequally spaced abscissas.

A second source of inaccuracy arises directly from
the behavior of the critical solution itself - as the
critical solution evolves each successive echo appears on
spatial and temporal scales a factor $e^{2 \Delta}
\approx 31$ finer than its predecessor. This problem is
solved as follows:  The number of grid points
decreases during the evolution. 
Once a null trajectory arrives at the origin
$r=0$ it bounces and disperses along $u =const.$ to
infinity. The grid point is therefore  lost when the 
light ray hits the origin. As the number of points decrease by 
half we double the number of grid points and we interpolate
the new grid points which are half way in between the old
ones. We chose the outermost grid point in such a way that
this  ingoing light ray 
hits the zero-mass singularity of the critical solution
itself. Thus the doubling of the grid gives us just the 
exact logarithmic scaling needed at the origin.

\section{DISCRETIZATION AND ERROR ANALYSIS}

Our grid is highly nonuniform in $u$ if a horizon forms -
as we approach to the horizon the step size $\Delta u$
decreases rapidly. Furthermore, even if the grid is evenly
spaced initially, it will not remain so during the
evolution \cite{Goldran}. However, when we consider two
grids in which one has twice the number of grid points
as the other, the coarser grid spacing will remain twice
the size of the finer grid, when we consider points at
the same physical location. We will discuss, therefore,
our numerical convergence in terms of the initial grid
spacing.

We denote the relative size of the initial grid spacing by
$l$. From the condition (\ref{eq24}) we see that the
spacing $\Delta u$ is proportional to $l$. As we have
mentioned, we treat Eqs. (\ref{eq16}) and (\ref{eq17}) as
ordinary differential equations in $u$, and we solve
these using a fifth-order Runge-kutta method
\cite{Press}, the error-term is $O(l^6)$.

The calculation of the quantities $\bar h, Q, g, \bar g, 
A_0$ and $M$ in these  equations is nontrivial, and is
given by Eqs. (\ref{eq7}), (\ref{eq8}), (\ref{eq13}),
(\ref{eq14}), (\ref{eq9}) and (\ref{eq19}) respectively.
The integrals are discretized using three-point Simpson
method, the error term in the integration is
$O(l^5f^{(4)})$, where $f^{(4)}$ is the fourth derivative
of the integrand $f$ \cite{Press}.

As we go from one grid to a finer one (i.e. doubling
the grid when the number of grid points reach half of
the original number) we have to interpolate to obtain
the values at the new grid points. The error term in the
interpolation is $O(l^4)$. 

As was mentioned in \cite{Chogolran}, the main source of
error is the boundary conditions $\bar g = g$  and 
$\bar h =  h$ at $r = 0$. We treat these 
boundary-conditions by approximating the true value of
$h(u,r=0)$ using an interpolation according to Eq.
(\ref{eq27}), the error term is $O(l^3)$. The situation is
even more complicated: The right hand-side of
(\ref{eq16}) contains an explicit factor of $1/r$, which
in the exact solution is canceled by the boundary
conditions. 

Three reasons lead to the crucial importance of checking
the behaviour of the solution near the origin, in order
to establish our confidence in the numerical results:

\begin{enumerate}

\item The risk of numerical instability caused by the
explicit factor of $1/r$.

\item The solution at some value of $r$ depends only on  
the solution at $r' <r$, so a numerical error in the
quantity $h$ near $r = 0$, would cause an error at each
$r$, and in all the relevant physical quantities.

\item The critical solution itself, which is the main
issue of this work, appears on ever smaller spatial
scales during its evolution, leading to the formation of
a zero-mass singularity at $r = 0$, and the importance of
our confidence in the numerical solution near the origin is
therefore clear.
\end{enumerate} 

As we can see, $h(u, r=0)$ is the basic quantity which
influence all the other physical quantities, and at each
and every value of $r$.

Figure 1 displays the error in $h(u, r=0)$. In this
figure the initial data is of family $(c)$, with
amplitude $A = 1.1$, and $e = 1$ (see V) for a
discussion of initial data). For these initial conditions
the gravitational field is strong $(A > A^*)$, and the
scalar-field undergo a terminal gravitational collapse
into a charged black-hole.

\begin{figure}[ht]
\centerline{\epsfig{file=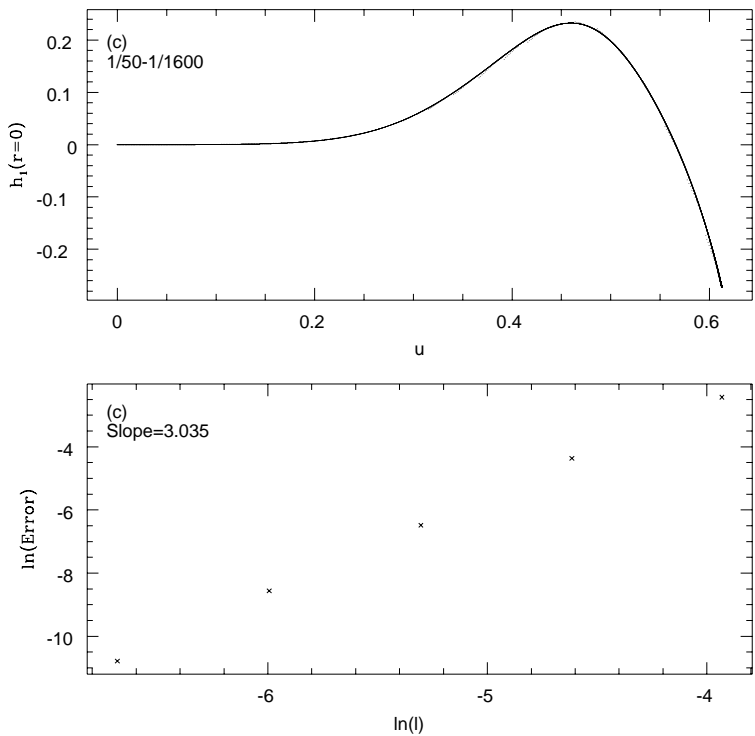,height=300pt}}
\caption{{\it The convergence of $h(u, r=0)$ with
decreasing grid size. The upper panel establishes
visually the convergence of the code, by showing the
real part of the scalar field $h_1(u, r=0)$ for
different relative grid spacing: 1/1600, 1/800, 1/400,
1/200, 1/100 and 1/50. The numerical errors are so small
that the different {\it six} curves (for the different
grids) actually overlap. The bottom panel establishes
quantitatively the stability and convergence of the
code. It displays the $\ln$ of the error (compared to a
reference solution with 1600 grid points) as a function
of $\ln$ of the initial grid size $l$. The slope is
$3.03 \pm 0.06$. This slope indicates  that the
error near the origin is largely caused by our handling
of the $r = 0$ boundary conditions. The initial data is
of family (c) with amplitude $(A > A^*)$ and $e = 1$.}}
\end{figure}

Due to the fact that there are no useful analytical
solutions available, we use of the numerical
solution itself as the reference solution. The reference
solution was taken as the solution with 1600
grid-points, and it was compared with the solutions of
800, 400, 200, 100  and 50 grid-points.

The difference of $h(u, r=0)$ between a given calculation
and the reference calculation was squared, summed (for
different vales of $u$) and the square root was taken.
The top part of Fig. 1 establishes visually that the code
converges: this graph displays the real part of the
scalar field $h_1(u,r=0)$ for different relative spacing
of the grid-points: 1/1600, 1/800, 1/400, 1/200, 1/100
and 1/50. The numerical errors are so small that the six
lines (for the different grids) actually overlap.

The bottom part of Fig. 1 establishes empirically the
stability and convergence of the code when we decrease
the grid spacing: the numerical error varies as $l^m$,
where $m = 3.03 \pm 0.06$. The fact that the power is
close to 3 at the origin is an indication that this error
is caused by our handling of the $r = 0$ boundary
conditions.

Furthermore, we have performed by a similar manner a
calculation of the error in $h(u, r = 0.03)$. The top
part of Fig. 2 displays again the fact that the
numerical errors are so small that the different lines
(for the different grids) actually overlap. The bottom
part of Fig. 2 establishes once again the stability and
convergence of the code: the error varies as $l^m$,
where $m = 4.01 \pm 0.05$. This value of the power
indicates that in this regime the discretization error
due to the interpolation which we use in order to
maintain the number of grid-points is dominant.

\begin{figure}[ht]
\centerline{\epsfig{file=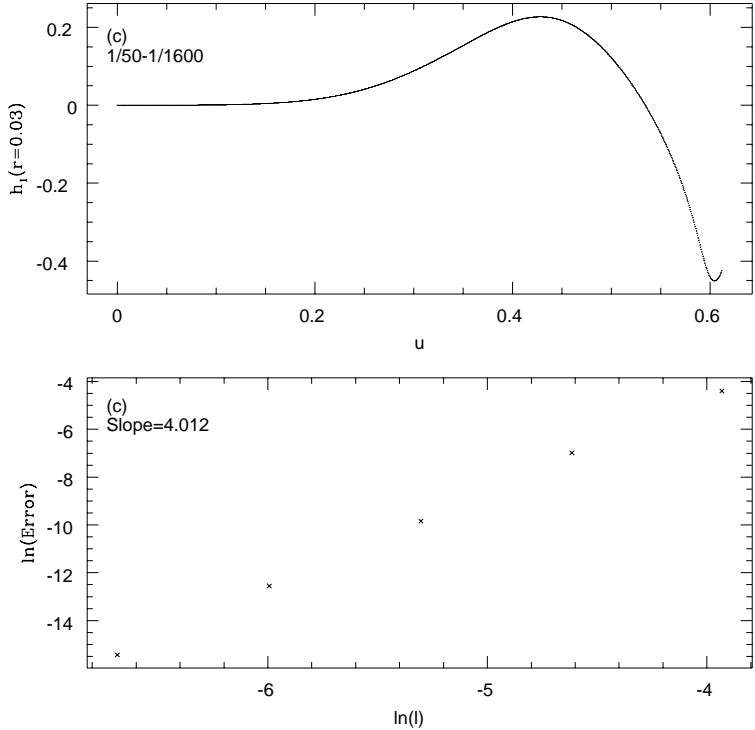,height=300pt}}
\caption{\it The convergence of $h(u, r = 0.03)$ with
decreasing grid size $l$. The upper panel establishes
visually the convergence of the code, by showing the real
part of the scalar field $h_1(u, r=0.03)$ for different
relative grid spacing: 1/1600, 1/800, 1/400, 1/200,
1/100 and 1/50. The numerical errors are so small that
the different {\it six} curves actually overlap. The
bottom panel establishes quantitatively the stability
and convergence of the code. It displays the ln of the
error (compared to a reference solution with 1600
grid points) as a function of ln of the initial grid
size, $l$. The slope is $4.01 \pm 0.05$. This slope
indicates that in this regime the discretization error
due to the interpolation is dominant. The initial data
is the same as for Fig. 1.}
\end{figure}

\section{THEORETICAL PREDICTIONS VS. NUMERICAL RESULTS}

In this section we present our theoretical predictions
and numerical results for the gravitational collapse of
a scalar field, for both the uncharged and for the
charged cases. The numerical results that we present
arise from a study of several families of solutions,
whose initial scalar field $\phi \equiv \bar h$ profiles
are listed in Table I:

\bigskip

\begin{center}
\begin{tabular}{|c|c|c|}\hline
Family & Form of initial data & \quad $e$ \quad \\
\hline   
(a) & $\phi (r) = Ar^2 \exp \left[ -
\left({{r - 0.2}\over{0.1}}\right)^2 \right]$ & \quad 0
\quad \\  
(b) & $\phi(r) = A \exp (-44 r) \cos (100 r)$ &
\quad 0 \quad \\
(c) & \quad $\phi(r) = A r^2 \exp \left[- \left(
{{r - 0.25}\over{0.1}}\right)^2 \right] + iA r^2
\exp \left[- \left( {{r - 0.15}\over{0.1}}\right)^2
\right]$ \quad & \quad 1 \quad \\  
(d) &  \quad $\phi (r) = A \exp (44r) \cos(100r) + iA \exp
(-75 r) \cos (200r)$ \quad & \quad 1 \quad \\
\hline
\end{tabular}
\end{center}

\bigskip

Families (a) and (b) represent uncharged scalar-field
while families (c) and (d) represent charged (complex)
scalar-field. The amplitude $A$ is the critical
parameter $p$. 

\subsection{THE CRITICAL SOLUTION}

In this section we discuss the critical solution $(p -
p^*)$ itself, both for the previously studied uncharged
case and for the newly studied charged one. First, we
had to find the value of the critical parameter. This
was done by a binary search until we found the value of
the critical parameter to the desired accuracy.

Let $u^*$ denote the value of $u$ at which the
singularity forms. We define
\begin{equation}\label{eq28}
T \equiv -\ln \left[ (u^* - u) /u^* \right]~;
\end{equation}
\begin{equation}\label{eq29}
R \equiv r/(u^* - u) = (r/u^*)e^T~.
\end{equation}
In terms of these variables, the critical solution for
the neutral case is characterized by discrete
self-similarity \cite{Choptuik}, i.e. $h(R,t)$ and other
form-invariant quantity such as $M/r$ or $dM/dr$ are
periodic functions of $T$. 

The generalization to the charged case is not
trivial. Charge-conjugation destroys the invariance of
the evolution-equations under the rescaling $u \to au, r
\to ar$, where $a$ is an arbitrary positive constant.
This invariance characterizes the system in the neutral
case and it is essential for the appearance of
critical, self similar behaviour. The lack of this
invariance raises the interesting question whether
critical behaviour will appear in the charged case.

Using a semi-qualitative argument, which is based upon
different behaviour of the mass and charge under the
rescaling $r \to ar$, we might be able to answer this
question.
As the critical solution evolves its structure appears
on ever smaller spatial (and temporal) scales. For the
special neutral case we know that each successive echo
appears on spatial and temporal scales which are a factor
$a^{-1} \equiv e^{2 \Delta} \approx 31$ smaller than its
predecessor. Under the rescaling $r \to ar$ we have:
\begin{eqnarray}
\bar h & \to & \bar h~;  \label{eq30}\\
g & \to & g~; \label{eq31}\\
Q & \to & a^2Q~; \label{eq32}\\
A_o & \to & aA_o~; \label{eq33}\\
\bar g \equiv \bar g_o + \bar g_e & \to & \bar g_o + a^2
\bar g_e~; \label{eq34}\\
M \equiv M_o + M_e & \to & aM_o + a^3 M_e~, \label{eq35}
\end{eqnarray}
where $\bar g_o, M_0$ are the parts of $\bar g$ and
$M$ which do not depend on $e$, and $\bar g_e, M_e$ are
the additions in the charged case.
Since  $a < 1$ we learn from Eqs.
(\ref{eq30}-\ref{eq35}) that as the field evolves, the
significance and influence of the charge on the evolution
near the origin is reduced.

Furthermore, as the oscilations proceed the scalar field approaches a
real function times a constant phase \cite{GundlachGarcia}. This leads
to an addtional decrease in the charge of the form:
\begin{eqnarray}
Q & \to & a^\xi  Q~; \label{eq32a}\\
A_o & \to & a^\xi A_o~; \label{eq33a}\\
\bar g \equiv \bar g_o + \bar g_e & \to & \bar g_o + a^{2\xi}
\bar g_e~; \label{eq34a}\\
M \equiv M_o + M_e & \to & aM_o + a^{2\xi} M_e ~, \label{eq35a}
\end{eqnarray}
where $\xi$ is a positive constant (see Section V.D).

When both effects are combined together  
$Q/M$ decreases  with each echo,
approximately from $Q/M \to a^(1+\xi) (Q/M)$. Looking at
the right hand side of equation (\ref{eq16}) we find that
the various terms scale according to
\begin{eqnarray}
{{1}\over{2r}} \left[ g - (\bar g_o + \bar g_e) \right]
(h - \bar h) & \to & {{1}\over{2ar}} (g - \bar g_o) (h -
\bar h) - {{a^{(1+2\xi)}}\over{2r}} \bar g_e (h - \bar h)~;
\label{eq36}\\ 
{{Q^2}\over{2r^3}} (h - \bar h)g & \to &
{{a^{(1+2\xi)} Q^2}\over{2r^3}} (h - \bar h)g~; \label{eq37}\\
{{eQ}\over{2r}}g\bar h & \to & {{a^{(1+\xi)} eQ}\over{2r}}~;
\label{eq38} \\ 
e h A_o & \to & a^{(1+\xi)} e h A_o~. \label{eq39}
\end{eqnarray} 
From here we learn that as the evolution proceeds the
last three terms in the right hand side of equation
(\ref{eq16}) becomes smaller relative to
the first term by a factor larger than  $a^2$ with each echo. From
these arguments we expect to find that for near-critical
evolutions the influence of the charge on the evolution
should decrease with each echo. 

We conjecture that in
the precisely critical case $(p = p^*)$ and in the limit
of an infinite train of echoes, the influence of the
charge on the evolution near origin is ``washed out" and
we expect to have the Choptuik solution. In practice,
using the fact that $a \ll 1$ we expect the influence of
the charge on the evolution to be negligible even after a
small number of echoes. Thus, once echoing begins the
solution will approach the neutral one rapidly. Our
numerical solution verifies these predictions. Fig. 3
shows the quantity max$(2M/r)$ as a function of $T$ for
near-critical evolution of families (a), (b), (c) and
(d). The solutions of families (b)-(d) were shifted
horizontally but not vertically with respect to family
(a) in order that the first echo of each family will
overlap the first echo of family (a). After an initial
phase of evolution the quantity max$(2M/r)$ settles down
to a periodic behaviour in $T$. The period is $\Delta
\approx 1.73$ which corresponds to previous numerical
results found by Choptuik \cite{Choptuik} for the neutral
case.

\begin{figure}[ht]
\centerline{\epsfig{file=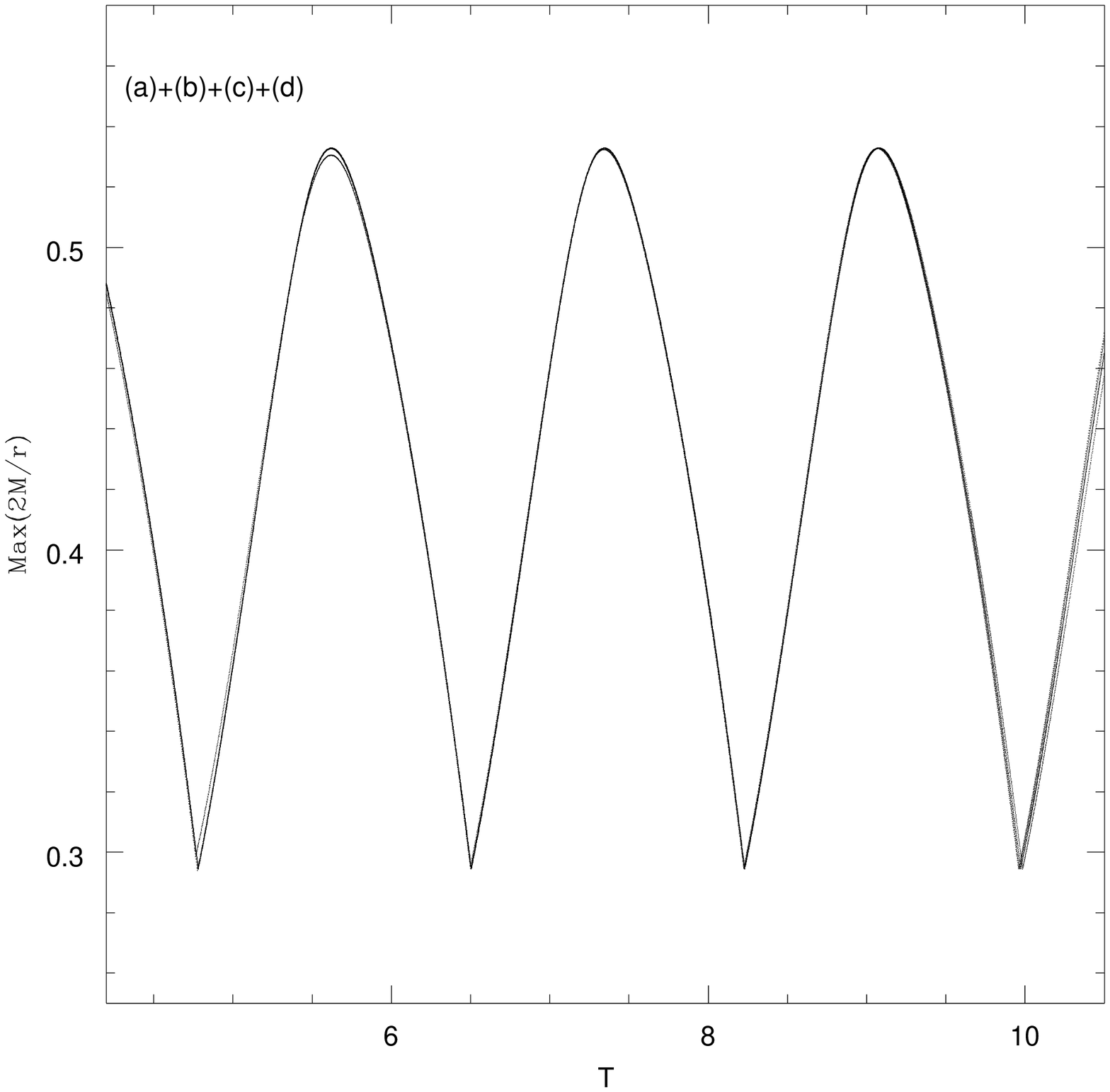,height=200pt}}
\caption{\it Illustration of the conjectures {\it
universality} of the critical evolution in the
gravitational collapse of a charged (complex) scalar
field. Max$(2M/r)$ is plotted as a function of the
logarithmic time $T$ for the neutral families (a), (b)
and for the {\it charged} families (c) and (d) (from
near-critical evolutions). The curves were shifted
horizontally (but not vertically) in order to overlap
the first echo of each family with the first one family
(a). After an initial evolution the quantity
Max$(2M/r)$ settles down to a behaviour which is
periodic in $T$. The period is $\Delta \approx 1.73$,
which corresponds to previous numerical results found by
Choptuik for the neutral case. This Figure provides a
numerical evidence for the conjectured generalization of
the echoing behaviour for the {\it charged} case.}
\end{figure}

Fig. 3 also provides a numerical evidence for the {\it
universality} of the strong-field evolution of a
critical configuration. In order to verify whether or
not this universality exist for each and every value $r$
we display in Fig. 4 profiles of $M/r$ as a function of
$R$ for $u$ values at which the quantity max$(2M/r)$
takes its maximum as a function of $T$. (It should be
emphasized that Fig. 4 is composed of {\it 16} profiles
- 4 for each family.) The close similarity of the
profiles illustrates two important properties:

\begin{enumerate}
\item For each family by itself, the critical solution
is periodic. Each successive echo appears on spatial and
temporal scales a factor $e^{2 \Delta} \approx 31$ finer
than its predecessor.
\item The close similarity of the profiles illustrates
the uniqueness of the critical solution, independent of
the initial profile and charge.

\begin{figure}[ht]
\centerline{\epsfig{file=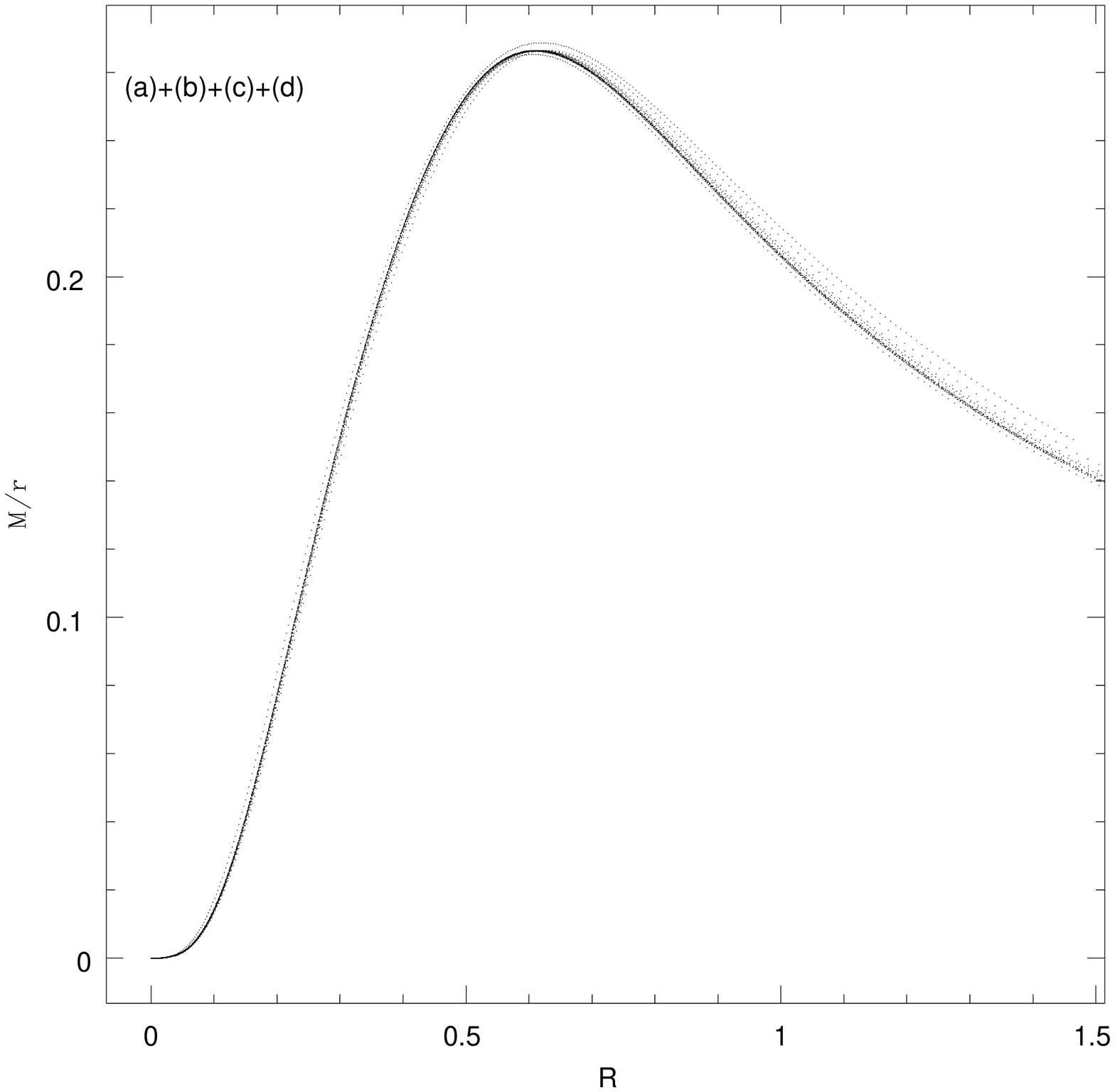,height=200pt}}
\caption{\it The profiles of $M/r$ for each of the
four families, are plotted as a function of the
logarithmic coordinate $R$ in those times at which the
quantity Max$(2M/r)$ reaches its maximum as a function
of $T$. (It should be emphasized that this figure is
composed from {\it 16} profiles -- 4 for each family.)
For each family by itself, each echo appears on spatial
and temporal scales a factor $e^{2\Delta} \approx 31$
finer than its predecessor. Furthermore, the close
similarity of the profiles illustrates the uniqueness
of the critical solution, independent of the critical
profile and charge.
}
\end{figure}

\end{enumerate}

\subsection{SCALING BEHAVIOR OF THE BLACK-HOLE MASS.}

We turn now to the power-law dependence of the black-hole
mass. The critical solution by itself does not yield the
black-hole scaling relation; we should perturb the
critical initial conditions, which has a
self-similar character. This would lead to dynamical
instability - a growing deviation from the critical
evolution towards either subcritical dissipation or
supercritical charged black-hole formation. To
describe the run-away from the critical evolution we
consider a perturbation mode with a power-law dependence
$\lambda (u^* - u)^{-\alpha}$ \cite{Coleman,Maison}, where
$\lambda \propto (p - p^*)$. Assume the range of validity of the
perturbation theory is restricted to some maximal
deviation $\sigma$ from the critical evolution, i.e. the
evolution is approximately self-similar until the
deviation from the critical solution reach the value
$\sigma$. From here on the evolution is outside the
scope of the perturbation theory -- there is subcritical
dissipation of the field or supercritical black-hole
formation. In either case, the evolution from this stage
on has no self-similar character.

We assume the perturbation in the initial conditions
develops into a charged black-hole the time $u_\sigma (p
- p^*)$ requires in order to have a $\sigma$ deviation
from the critical solution is given by the relation
\begin{equation}\label{eq40}
\lambda(u^* - u_\sigma)^{- \alpha} = \sigma~.
\end{equation}
Of course, a bigger initial perturbation requires a
smaller time $u_\sigma$ -- the horizon is formed
sooner. The logarithmic time $T_\sigma$ is given by
\begin{equation}\label{eq41}
T_\sigma = - {{1}\over{\alpha}} \ln (p - p^*) + b_k~,
\end{equation}
where $b_k$ depends on $\sigma$ and $u^*$.

In a following paper  \cite{Hod,HodPiran}
we prove that in order to have the
total logarithmic time $T_{BH}$ until the horizon
formation one should add a {\it periodic} term $F[ \ln
(p - p^*)]$ with a universal period, $\varpi$, which depends on
the previous universal parameters according to: $\varpi =
\Delta / \beta$, so that the full dependence of $T_{BH}$
on critical separation $p - p^*$, for the neutral case as
well as for the charged case, is given by
\begin{equation}\label{eq42}
T_{BH} = - {{1}\over{\alpha}} \ln (p - p^*) + F [ \ln (p -
p^*)] + b_k~. 
\end{equation}

Fig. 5 displays $-T_{BH}$ as a function of $\ln (a)$,
where $a = (A - A^*)/A^*$, for the neutral families (a),
(b) and for the {\it charged} families (c), (d) as
well. The points are well fit by a straight line whose
slope is $1/\alpha \approx 0.37$. This is consistent
with the relation (\ref{eq42}).
\begin{figure}[ht]
\centerline{\epsfig{file=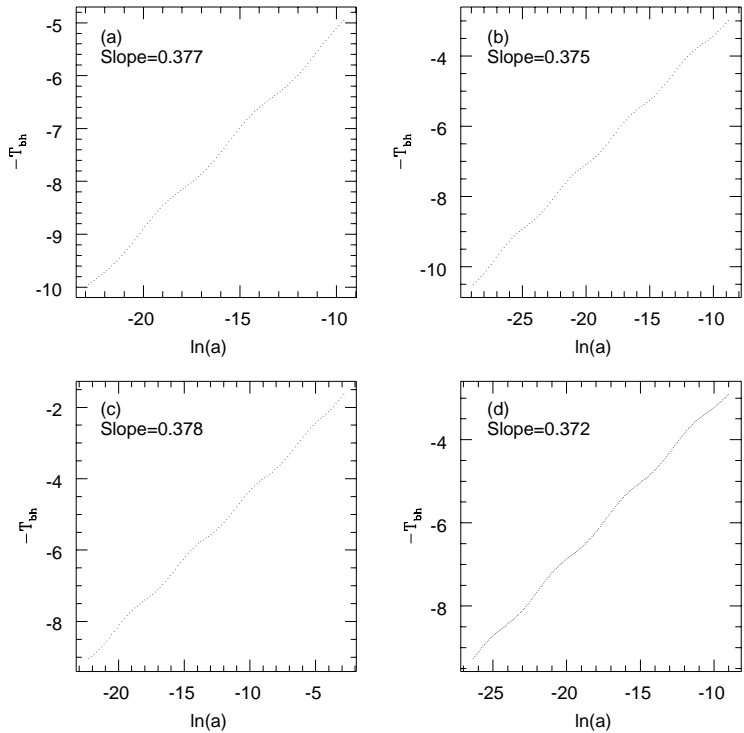,height=300pt}}
\caption{\it The logarithmic time of a black-hole
formation, plotted as $-T_{BH}$ as a function of
$\ln(a)$ (where $a = (p - p^*)/p^*$) for the four
families. The points are well fit by a straight line
whose slope is $1/\alpha \approx 0.37$. This is
consistent with the relation (\ref{eq41}). 
The oscillations
seen above the straight line are physical [1].
}

\end{figure}

The exponent $\alpha$ is related to $\beta$
the critical exponent that describes the
power-law dependence of the black-hole mass.
We define $M^{(n)}$ as the mass after $n$ echoes.
Following this definition we define
\begin{equation}\label{eq43}
M^{(o)} = M_o^{(o)} + M_e^{(o)} \equiv (1 + C)M_o^{(o)}~,
\end{equation} 
where $M_o$ is the part of $M$ which does not depend on
$e$ and $M_e$ is the charge contribution in the charged case.
From (\ref{eq35} and \ref{eq35a})   it follows that
\begin{equation}\label{eq44}
M^{(n)} = M_o^{(o)}e^{-n\Delta} + C M_o^{(o)}e^{-(3+2\xi) n \Delta}~. 
\end{equation}
We substitute (\ref{eq41}) into (\ref{eq44}) and obtain
\begin{equation}\label{eq45}
\ln M_{BH} = c_k + \beta \ln (p - p^*) + O[e^2(p - p^*)^{2
\beta(1 + \xi)}]~, 
\end{equation}
and find that $\beta = 1/ \alpha$ and $c_k$ is a
family-dependent constant. Here we have assumed that
$M^{(o)}$ and $C$ do not depend on $p - p^*$.

Thus, according to the perturbation theory, the
critical exponent $\beta$ describes both the
system's response to perturbations to the critical
evolution, and the deviation rate from it. 
In a following paper \cite{HodPiran,Hod} \footnote{see also
\cite{Gundlach2}} we show that one should add a periodic
term  $\Psi [\ln (p - p^*)]$  with a universal
period $\varpi = \Delta / \beta$ to the scaling relation
(\ref{eq45}).

Fig. 6 displays $\ln (m)$ as a function of $\ln (a)$
for the neutral families (a), (b) and for the {\it
charged} families (c) and (d) as well, where $m$ is the
normalized black-hole mass in units of the initial mass
in the critical solution. The points are well fit by a
straight line whose slope is $\beta\approx 0.37$. This
is consistent with the prediction according to which
$\beta = 1/\alpha$.
The measured value of the critical parameter $\beta$ in
the charged case agrees with the one of Choptuik for the
special non-charged case. Thus, Fig. 6 presents a
generalization of the mass-scaling phenomenon for the
{\it charged} case.

\begin{figure}[ht]
\centerline{\epsfig{file=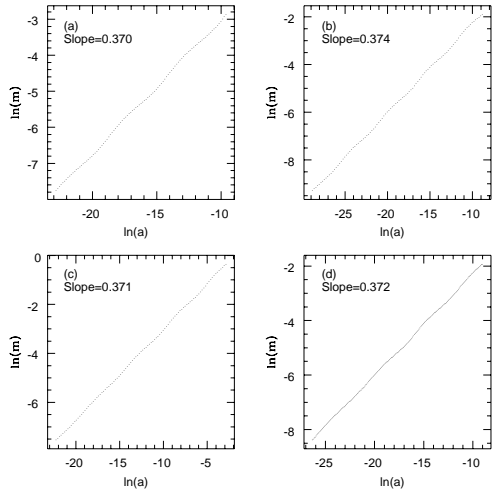,height=200pt}}
\caption{\it Scaling of the black-hole mass.
$\ln(m)$ is plotted vs. $\ln (a)$ for the neutral
families (a), (b) and for the {\it charged} families
(c) and (d), where $m$ is the normalized black-hole
mass in units of the initial-mass in the critical
solution. The points are well fit by a straight line
whose slope is $\beta \approx 0.37$. This is
consistent with the predicted relation $\beta =
1/\alpha$. The measured value of the critical exponent
$\beta$ in the charged case agrees with the one
previously found for the special neutral case. Thus,
this figure presents a generalization of the
mass-scaling phenomena for the {\it charged} case.
The oscillations above the straight line are physical 
[1,14,15].}
\end{figure}

In the neutral case the black-hole mass is proportional to
its radius and either the mass or the radius could be
the physical order-parameter.
In the charged case the black hole radius equals $M + (M^2 -
Q^2)^{1/2}$ and in general it is no longer proportional
to $M$.  However, as we approach the
critical-solution, the significance of the charge 
decreases with each echo. The quantity $Q/M$ becomes
smaller as the black-hole mass gets smaller and it is
impossible to distinguish between the two possibilities.

\subsection{EXTERNAL PERTURBATIONS TO THE CRITICAL
SOLUTION}

So far, the mass-scaling relation of the black-hole, and
the influence of perturbations on the critical evolution
itself, have been studied in context of internal
perturbations in the initial conditions, such as a
deviation of the field's amplitude from the critical
one. It is of interest to study the behaviour of the
critical solution under {\it external} perturbations. In
particular, we have studied the behaviour of the
critical evolution under two external perturbations:

\begin{enumerate}
\item Addition of a cosmological-constant $\Lambda$ to
the critical solution itself.
\item An addition of a charge-conjugation $e$ to
critical initial-conditions of a complex neutral
scalar-field. 
\end{enumerate} 

Adding an external perturbation to the critical
solution, which has a self-similar character, is
expected to yield dynamical instability -- a growing
deviation from the critical evolution toward either
subcritical dissipation or supercritical black-hole
formation. In the situation where the external
perturbation to the critical solution (which, in the
unperturbed case, is expected to form a zero-mass
singularity) leads to a formation of a black-hole, we
have studied the question whether or not there is a
power-law dependence of the black-hole mass on the {\it
external} parameter: $\Lambda$ or $e$.

\subsubsection{Addition of a cosmological constant
$\Lambda$}

The amplitude was first set to its critical value $A^*$
(for $\Lambda = 0$). We then add a cosmological-constant,
$\Lambda$, and study its influence on the evolution of
the critical solution. $\Lambda > 0$ led to a
dissipation of the critical solution and, on the other
hand, $\Lambda < 0$ was found to give a finite-mass
black-hole formation. 

We have studied the behaviour of the black-hole mass as
a function of the cosmological constant. This is
analogue to the addition of an {\it external} magnetic
field to a system of magnetic moments and studying the
magnetization dependence on the strength of the external
field exactly at the critical temperature $T_c$ (at $t =
T_c$ the magnetization is zero without an external
magnetic field, just as the zero-mass singularity for $A
= A^*$ and $\Lambda = 0$). The analogy arises from the
fact that in both cases the {\it external} perturbation,
magnetic-field or a cosmological-constant, forces the
order-parameter to have a non-zero value at $T = T_c$ or
at $A = A^*$, correspondingly. 

For $\Lambda \not= 0$, the generalization of Eq.
(\ref{eq14}) is 
\begin{equation}\label{eq46}
\bar g (u, r) = 1/r \int\limits_a^r (1 -
(Q^2/r^2) - \Lambda r^2) gdr~. 
\end{equation}
In addition, one should add the term 
\begin{equation}\label{eq47}
- \Lambda/2rg(h -
\bar h)~,
\end{equation} 
in the right-hand side of Eq. (\ref{eq16}). Evidence
for the {\it universality} and the power-law dependence
of the black hole mass on the {\it external} parameter
$\Lambda$ is shown in Fig. 7 which displays $\ln (m)$ as
a function of $\ln (|\Lambda|)$ for critical
initial-conditions $(A = A^*)$. The points are well fit
by a straight line, i.e.
\begin{equation}\label{eq48}
M_{BH} \propto |\Lambda|^{1/\delta}~,
\end{equation}
where $1/\delta \approx \beta$.

\begin{figure}[ht]
\centerline{\epsfig{file=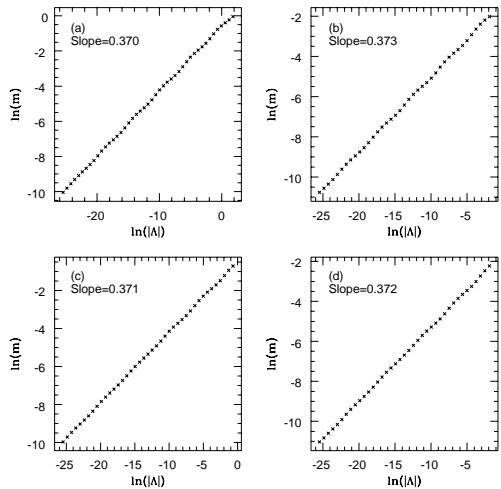,height=200pt}}
\caption{\it Power-law dependence of the black-hole
mass on the cosmological constant. $\ln (m)$ is
plotted vs. $\ln (|\Lambda|)$ for the four families. The
points are well fit by a straight line whose slope is
$1/\delta$, where $1/\delta \approx \beta$. This
provides an evidence for the power-law dependence of
the black-hole mass on {\it external} parameters. The
initial conditions are for $A = A^*$.}
\end{figure}

\subsubsection{Addition of a charge-conjugation $e$}

The amplitude was first set to its critical value $A^*$
for the families (c) and (d), with $e = 0$. For $e = 0$,
the initial conditions represent uncharged complex
scalar-field. We then add a charge-conjugation, $e$, and
studied its influence on the evolution of the critical
solution. We have studied the behaviour of the
black-hole mass as a function of the charge-conjugation
$e$ (keeping the amplitude on its critical value $A^*$).
The numerical results are shown in Fig. 8, which
displays $\ln (m)$ as a function of $\ln (e^2)$.
The points are well fit by a straight line, i.e.
\begin{equation}\label{eq49}
M_{BH} \propto |e|^\gamma~,
\end{equation}
where $\gamma \approx 2\beta$.

\begin{figure}[ht]
\centerline{\epsfig{file=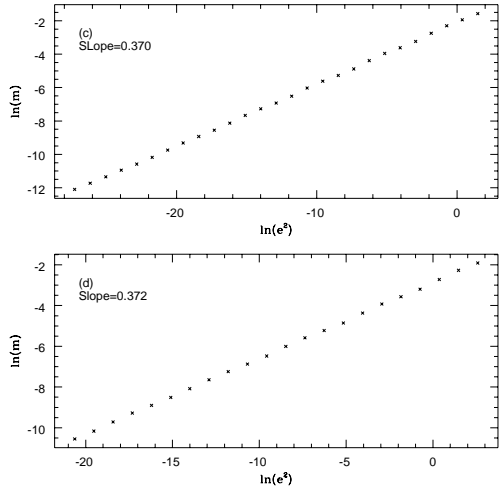,height=200pt}}
\caption{\it Power-law dependence of the black-hole
mass on charge-conjugation $e$. $\ln (m)$ is plotted
vs. $\ln (e^2)$ for the charged families (c) and (d).
The points are well fit by a straight line whose slope
is $\gamma/2 \approx \beta$. The initial conditions
are for $A = A^*$.}
\end{figure}

In order to have a more detailed picture of the
formation for a charged black-hole near the
phase-transition we have studied the conditions
required the formation of a charged black-hole and
the dependence of its mass on the critical separation $|p
- p^*|$ and on the charge-conjugation $e$. This was done
both for subcritical initial-conditions and for
supercritical initial conditions as well, which are
close to the phase-transition $(|A - A^*| \ll A^*)$.

Fig. 9 displays the dependence of the black-hole mass
which forms from subcritical initial conditions $(A
< A^*)$ as a function of $\ln (a)$ and $\ln (e)$, for
family (c). This provides a numerical evidence for the
conjecture that external perturbation such as
charge-conjugation can lead to a black-hole formation with
subcritical initial-conditions $(A < A^*)$. Charge
oppose gravitation but the electromagnetic energy density
also contributes to the gravitational binding, and by
doing so it permits the formation of a black-hole even
from subcritical initial-conditions.

\begin{figure}[ht]
\centerline{\epsfig{file=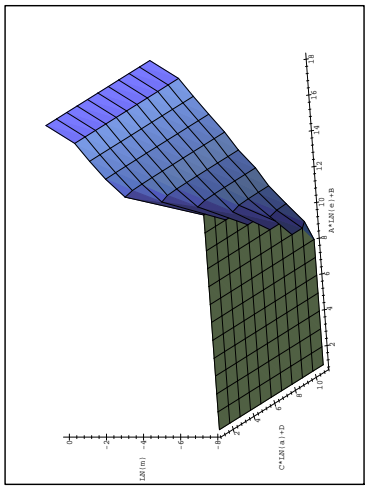,height=200pt}}
\caption{\it The black-hole mass, $\ln(m)$, is plotted
as a function of $\ln(a) = \ln(A^* - A)/A^*$ and
$\ln(e)$, for subcritical initial conditions $(A < A^*)$.
The charge opposes gravitation but the electromagnetic
energy also contributes to the gravitational binding, and
by doing so it permits the formation of a black-hole even
from subcritical initial conditions. Black holes do not
form in the flat regime. A, B, C and D are normalization
constants: A = 1, B = 16, C = -1, D =-11.061}
\end{figure}

The critical charge-conjugation $e^*$, needed to obtain
large enough deviations from the otherwise evolution,
followed by black-hole formation from subcritical initial
conditions, is well described by a power-law dependence
on critical-separation $|A - A^*|$:
\begin{equation}\label{eq50}
e^*\propto(p^* - p)^\epsilon ~,
\end{equation}
where $\epsilon = 0.5$ (see Fig. 10).

\begin{figure}[ht]
\centerline{\epsfig{file=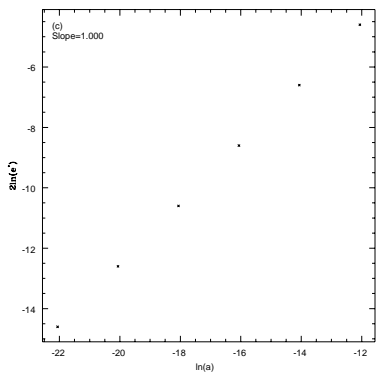,height=150pt}}
\caption{\it Power-law dependence of the critical
charge-conjugation $e^*$ on the critical separation $|A -
A^*|$, for subcritical $(A < A^*)$ initial conditions.
$e^*$ is the minimal charge-conjugation needed for
black-hole formation from subcritical
initial-conditions.}
\end{figure}

Fig. 11 displays the dependence of the black-hole mass
formed from supercritical initial conditions
$(A > A^*)$ as a function of $\ln (a)$ and $\ln (e)$, for
family (c). We learn that near the phase-transition, the
larger the critical separation $p - p^*$ and the
charge-conjugation $e$, the larger is the black-hole
mass. The slope of the right edge of the surface is
$\beta$, which is consistent with Eq. (\ref{eq45}). The
slope of the left edge of the surface is $2 \beta$, which
is consistent with Eq. (\ref{eq49}).

\begin{figure}[ht]
\centerline{\epsfig{file=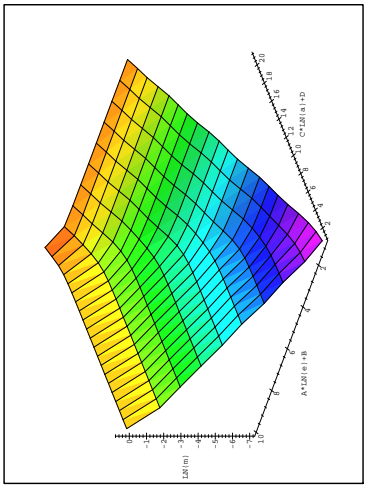,height=200pt}}
\caption{\it The black-hole mass, $\ln(m)$, is plotted
as a function of $\ln(a) = \ln(A - A^*)/A^*$ and
$\ln(e)$, for supercritical initial conditions $(A >
A^*)$. Near the phase-transition, the larger the critical
separation $ p - p^*$ and the charge-conjugation $e$, the
larger is the black-hole mass. The slope of the right
edge of the surface is $\beta$, which is consistent with
Eq. (\ref{eq45}). The slope of the left edge of the
surface is $2\beta$, which is consistent with Eq.
(\ref{eq49}). A, B, C, D are normalization constants: A =
1, B = 8, C = 1, D = 23.061.
}
\end{figure}

It should be emphasized that the tight relation between
the various critical-exponents $\beta, \delta$ and
$\gamma$, which all describe the {\it instability} of
the critical evolution under a variety of different {\it
perturbations}, is a strong evidence supporting the
conjecture that there exists one mechanism
which can explain the power-law dependence of the
black-hole mass on the various parameters, both for
internal perturbations in the initial conditions, and
for external perturbations as well. 

\subsection{THE CHARGE-MASS RELATION FOR BLACK HOLES}

We have shown in 5.A that  the black-hole charge tends
to zero faster than its mass for $p - p^* \to 0$. We demonstrate 
now that $Q \rightarrow 0$ as a power law with a critical exponent
larger than 
$2 \beta$ (see also \cite{GundlachGarcia} for an independent analysis
of this problem)..  Define
$Q^{(n)}$ as the charge after $n$ echoes. From (\ref{eq32} and \ref{eq32a})
it follows that
\begin{equation}\label{eq51}
Q^{(n)} = Q^{(0)}e^{-(2+\xi) n\Delta}~.
\end{equation}
Substituting (\ref{eq41}) into (\ref{eq51}) and assuming
that $Q^{(0)}$ does not depend on $p - p^*$ we obtain
\begin{equation}\label{eq52}
\ln |Q_{BH}| = (2+\xi) \beta \ln (p - p^*) + d_k~,
\end{equation}
where $d_k$ is a family-dependent constant.

From this analytic argument we deduce two conclusions:

\begin{enumerate}
\item The black-hole charge is expected to have a
power-law dependence on critical separation $p - p^*$,
where the critical-exponent $\eta$ is closely related to
the critical-exponent $\beta$ of the mass according to
$\eta = (2+\xi) \beta$.
\item The black-hole charge tends to zero with $p - p^*$
more rapidly than its mass. 
\end{enumerate}

Fig. 12 displays $\ln q_{BH}$ as a function of $\ln (a)$ for
near-critical black-holes, where $q_{BH}$ is the normalized black-hole
charge in units of the initial-charge in the critical-solution.  The
points are well fit by a straight line whose slope is $\eta\approx
0.88 $.  This is consistent with the relation (\ref{eq52}).

\begin{figure}[t]
\centerline{\epsfig{file=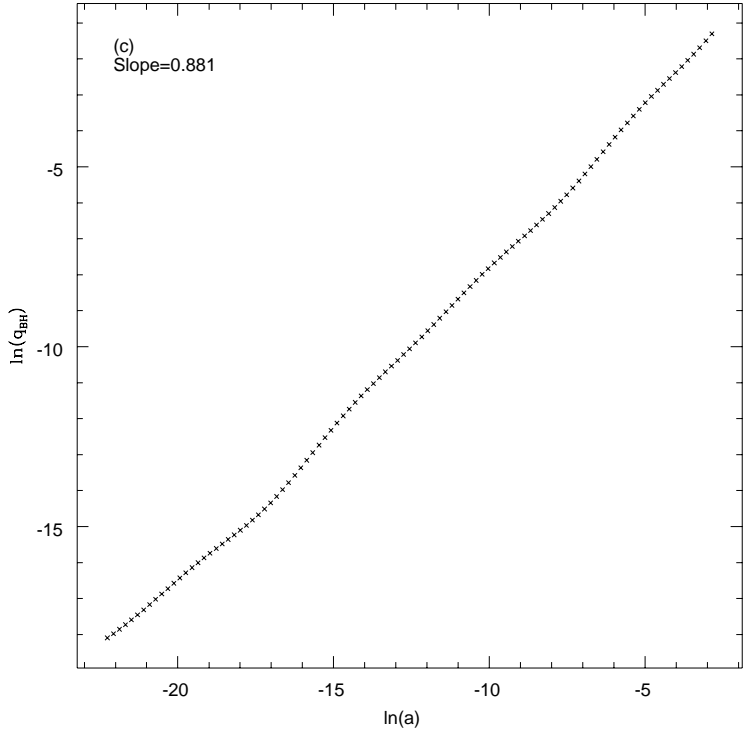,height=300pt}}
\caption{\it  Illustration of the conjectured
charge-scaling relation \ref{eq52}. $\ln(q_{BH})$ is plotted
vs. $\ln(a)$ for near-critical black-holes, where $q_{BH}$ is the
normalized black-hole charge in units of the initial-charge in the
critical solution. Data from the charged family (c) is shown. The
points are well fit by a straight line whose slope $\eta$ obeys the
relation $\eta > 2 \beta$. Thus for $p - p^* \to 0$ the black-hole
charge tends to zero more rapidly than its mass.  }
\end{figure}

Another way to determine the value of $\eta$ is directly from
Eqs. (\ref{eq51} and \ref{eq52}). Fig. 13 displays $\ln q^{(n)}$ as a
function of $n$, the number of echoes, along the critical
solution. The slope is $-(2+\xi) \Delta \approx -4.133$. Using $\Delta
\approx 1.73$ and $\beta \approx 0.37$ we find $\eta \approx 0.883$,
in agreement with the prediction, $\eta = 0.883 \pm 0.007 $, 
of Gundlach and Martin-Garcia\cite{GundlachGarcia} .
\begin{figure}[t]
\centerline{\epsfig{file=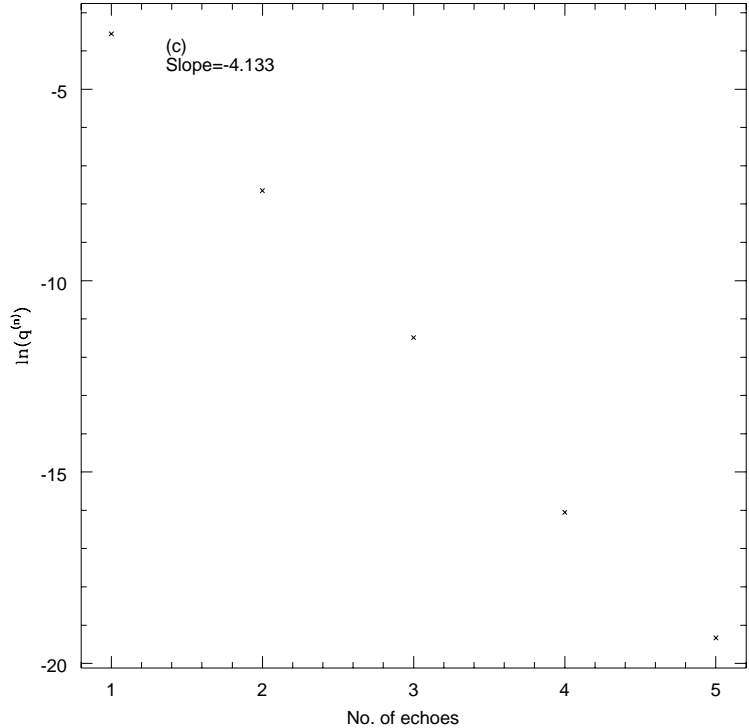,height=300pt}}
\caption{\it   The decrease of the charge with each echo
during the critical evolution. The points are will fit by a straight
line whose slope is $-(2 + \xi) \Delta \approx -4.133$.}
\end{figure}

The data shown in Figs. 12 and 13 come from the charged family
(c). For family (d) we have found that in general the
charge increases as the critical separation $ p - p^*$
increases although this correlation is not well
described by a power-law dependence. The reason for this
situation probably comes from the initial stage of
the evolution (note that the initial data contains layers of
both positive and negative charge whose relative magnitude depends
on $p$. Consequently, the assumption that $Q^{(0)}$ does not
depend on $p - p^*$ breaks down for this configuration.

Despite of this, our numerical results qualitatively
confirm our prediction -- the black-hole charge tends to
zero with $p - p^*$ faster than its mass. From here
we conclude black-holes with infinitesimal mass, which,
according to this mechanism, can be created from
near-critical evolutions, are neutral, or obey the
relation $Q_{BH} \ll M_{BH}$.

\section{SUMMARY AND CONCLUSION}

We have studied the spherical gravitational collapse of
a charged (complex) scalar-field. The main issue
considered is the generalization of the critical
behaviour, originally discovered by Choptuik for neutral
fields, for the general {\it charged} case. This
generalization is not trivial, since the introduction of a
charge-conjugation destroys the invariance of the
evolution equations under the rescaling $u \to au, r \to
ar$, which  characterizes the system in the 
neutral case and is crucial for the self-similarity of
the critical evolution. This lack of invariance hints that
charge-conjugation might destroy the phenomena in the
general situation. 

However, we have shown that the
significance and influence of the charge decreases
during the evolution and consequently the critical
behaviour appears. 
As $p - p^* \to 0$ the black-hole charge tends to zero faster than its
mass, i.e. $Q_{BH}/M_{BH}
\mathrel{\mathop{\kern0pt\longrightarrow}\limits_{p\to
p^*}} 0$. Thus, we conjecture that black-holes of infinitesimal mass,
which can be created from near-critical evolutions, are neutral, or
obey the relation $Q_{BH} \ll M_{BH}$. Consequently, we find both the
mass scaling relations for supercritical solutions and the echoing
phenomenon for the critical solution. The charge of the black hole
also depends, as a power law on the separation from criticality. The
critical exponent, is $\eta> 2 \beta$.

We have also studied the response of the critical
solution to {\it external} perturbations. In
particular, we have studied the behaviour of the critical
evolution under an addition of a cosmological-constant
$\Lambda$, and an addition of charge-conjugation $e$ to
critical initial-conditions of a complex neutral
scalar-field. We have found a power-law dependence 
of the black-hole mass on {\it external} parameters $\Lambda$
and $e$.

Under the influence of these external perturbations it is
possible to form black-holes from subcritical $(p <
p^*)$ initial conditions.
Once more, the  critical charge-conjugation $e^*$, which
is needed  to form  black-holes  from
subcritical initial-conditions, is well described by a
power-law dependence on critical separation $p^* - p$.

\acknowledgments
We thank Amos Ori for helpful discussions and Carsten Gundlach for
critical remarks.  This research was supported by a grant from the
US-Israel BSF and a grant from the Israeli Ministry of Science.

\end{document}